# The BiPublishers ranking: Main results and methodological problems when constructing rankings of academic publishers

Título en español: **El ranking BiPublishers: Principales resultados y problemas metodológicos en la contrucción de rankings de editoriales académicas**

Daniel Torres-Salinas[1], Nicolás Robinson-Garcia[2], Evaristo Jiménez-Contreras[3] and Enrique de la Fuente[4]

[1] *torressalinas@gmail.com*

EC3metrics spin-off, Universidad de Navarra, Granada

[2] *elrobin@ugr.es* [3] *evaristo@ugr.es*

EC3metrics spin-off and EC3 Research Group, Universidad de Granada, Granada

[4] *efg_39@ugr.es*

EC3metrics spin-off, Universidad de Granada, Granada

**Abstract**

We present the results of the Bibliometric Indicators for Publishers project (also known as BiPublishers). This project represents the first attempt to systematically develop bibliometric publisher rankings. The data for this project was derived from the Book Citation Index, and the study time period was 2009-2013. We have developed 42 rankings: 4 for by fields and 38 by disciplines. We display six indicators by publisher divided into three types: output, impact and publisher's profile. The aim is to capture different characteristics of the research performance of publishers. 254 publishers were processed and classified according to publisher type: commercial publishers and university presses. We present the main publishers by fields. Then, we discuss the main challenges presented when developing this type of tools. The BiPublishers ranking is an on-going project which aims to develop and explore new data sources and indicators to better capture and define the research impact of publishers.

**Resumen**

Presentamos los resultados del proyecto Bibliometric Indicators for Publishers (BiPublishers). Es el primer proyecto desarrolla de manera sistemática rankings bibliométricos de editoriales. La fuente de datos empeleada es el Book Citation Index y el periodo de análisis 2009-2013. Se presentan 42 rankings: 4 por áreas y 38 por disciplinas. Mostramos seis indicadores por editorial divididos según su tipología: producción, impacto y características editoriales. Se procesaron 254 editoriales y se clasificación según el tipo: comerciales y universitarias. Se presentan las principales editoriales por áreas. Después, se discuten los principales retos a superar en el desarrollo de este tipo de herramientas. El rankings Bipublishers es un proyecto en desarrollo que persigue analizar y explorar nuevas fuentes de datos e indicadores para captar y definir el impacto de las editoriales académicas.

**Palabras clave:** editoriales académicas; rankings; indicadores bibliométricos; Book Citation Index; editoriales universitarias

# Introduction

In the last years many advances have been made on the development of bibliometric databases including books and book chapters. These document types have been historically neglected from bibliometric analysis (Nederhof, 2006), however, the launch of products such as Google Scholar, Google Books, the Book Citation Index or their inclusion in databases such as Scopus has opened a wide scope of opportunities for their analysis (Kousha et al., 2011; Torres-Salinas et al., 2014). Similarly to journal rankings, one first step for including books and book chapters in the bibliometric toolbox may be to develop publisher rankings. There are already some initiatives following this line of thought (see for example Research School for Socio-Economic and Natural Sciences of the Environment, 2009). In a previous paper we suggested the development of academic publisher rankings (Torres-Salinas, et al., 2012) based on the Book Citation Index. This paper builds on the idea of developing academic rankings based on the Book Citation Index (Torres-Salinas et al., 2012). Here we present the results of the *BiPublishers-Bibliometric Indicators for Publishers* project (Robinson-Garcia et al., 2014) available at http://bipublishers.es. This is an initiative aimed at developing new methodologies and indicators that can better capture and define the research impact of academic and scholarly book publishers. It is an on-going initiative in which data sources and i ndicators are tested. Hence, the information displayed should not be used for research evaluation purposes. We consider academic publishers as an analogy of journals, focusing on them as the unit of analysis; an approach already suggested elsewhere (i.e., Torres-Salinas & Moed, 2009). We include six indicators for more than 100 publishers in four broad fields and 38 different disciplines. The data is based on the Thomson Reuters' Book Citation Index.

# Material and methods

## General description of the database used: The Book Citation Index

The Book Citation Index (BKCI) was released in 2011 aiming to shed light on the research performance of monographs. It filled a gap which was already noted by Garfield (1996), creator of the original Science Citation Index. The Thomson Reuters' Book Citation Index (BKCI) was launched in 2011. It provides large sets of citation and publication data on monographs and book chapters and it is included in the Web of Science Core Collection within the Web of Knowledge platform.

It covers scientific literature since 1999 and, as it occurs with the Science Citation Index, Social Sciences Citation Index and Arts & Humanities Citation Index, it follows a rigorous selection process using the following principle criteria (Testa, 2010): 1) currency of publications, 2) complete bibliographic information for all cited references, and 3) the implementation of a peer review process. As a recent product, the BKCI has important limitations that must be considered when analysing the results shown. Here we summarize the main ones (Torres-Salinas et al., 2014):

- *Language bias*. It is strongly biased towards English language speaking countries, as to date (November, 2014), 97.7% of the records are in this language.
- *Great concentration of publishers*. Only three publishers (Springer, Palgrave and Routledge) represent half of the database.
- *Dispersion of citations*. Due to the distinction between books and book chapters, citations to each of them are also considered as independent.

## Data processing and normalization

All results shown are based on the web version of the BKCI back in April 2014. The time period covered is 2009-2013. For this period 482,470 records where retrieved, distributed in 14 different document types (see Figure 2 from Robinson-Garcia et al., 2014). Regarding the construction of fields, this was made through the aggregation of Web of Science subject categories as presented in the BKCI. Unlike to what occurs with journals, books are individually assigned to one or more categories, meaning that a single publisher may have (and usually has) their output distributed among different categories. The aggregation of subject categories for fields and disciplines is available at http://bipublishers.es/wp-content/uploads/2014/10/5.FieldsandDisciplinesConstruction.xlsx

For each record we processed the bibliographic fields. The field Publisher was processed separately and normalised manually. We identified 342 different publishers although 254 were finally processed. In order to ensure reliable results, publishers had to meet at least one of the following criteria to be included in a ranking: a) have a minimum of five books indexed during the study time period; or b) have a minimum of 50 book chapters indexed during the study time period. In the normalisation process we adopted as a criterion that if a publisher had been acquired by another one, then all its output will be assigned to the latter one. Also, we assigned publisher types, differentiating between two types: 1) commercial and academic publishers, and 2) university presses.

**Table I.** Definition of the indicators displayed by publisher

| | *Output indicators* | |
|---|---|---|
| PBK | Total number of books | Total number of books published by a given publisher in a certain field or discipline for the study time period (2009-2013). Minimum threshold |
| PCH | Total number of book chapters | Total number of book chapters published by a given publisher in a certain field or discipline for the study time period (2009-2013) |
| | *Impact indicators* | |
| CIT | Total number of citations | Total number of citations received by a given publisher in a certain field or discipline. |
| FNCS | Field normalized citation score | Field Normalized Citation Score. Normalized citations received according to the normalized indicator as defined by Moed et al. (1995). |
| | *Publisher's profile* | |
| AI | Activity Index | Distribution of books in a given field or discipline according to the overall output of a given publisher and in reference to the distribution of the whole BKCI |
| ED | Percentage of edited items | Share of book chapters which belong to edited books from the total number of book chapters published by a given publisher in a certain field or discipline for the study time |

|  |  | period (2009-2013). |
|---|---|---|

## Results

### Brief description of the indicators and web platform

Table I shows the six indicators displayed for each publisher. As observed, three types of indicators were selected in order to capture different aspects of the research performance of publishers. The first type of indicators shows the output of the publisher (PBK and PCH). The second group focuses on impact indicators, including the raw number of citations received (CIT) and a normalized impact indicator (FNCS). Finally, the third type of indicators intends to characterize the publisher. In this case we have included the activity index (AI) and the share of edited chapters from their total output in a given field (ED).

**Figure 1.** Snapshot of the ranking for publishers in the discipline of Information Science & Library Science

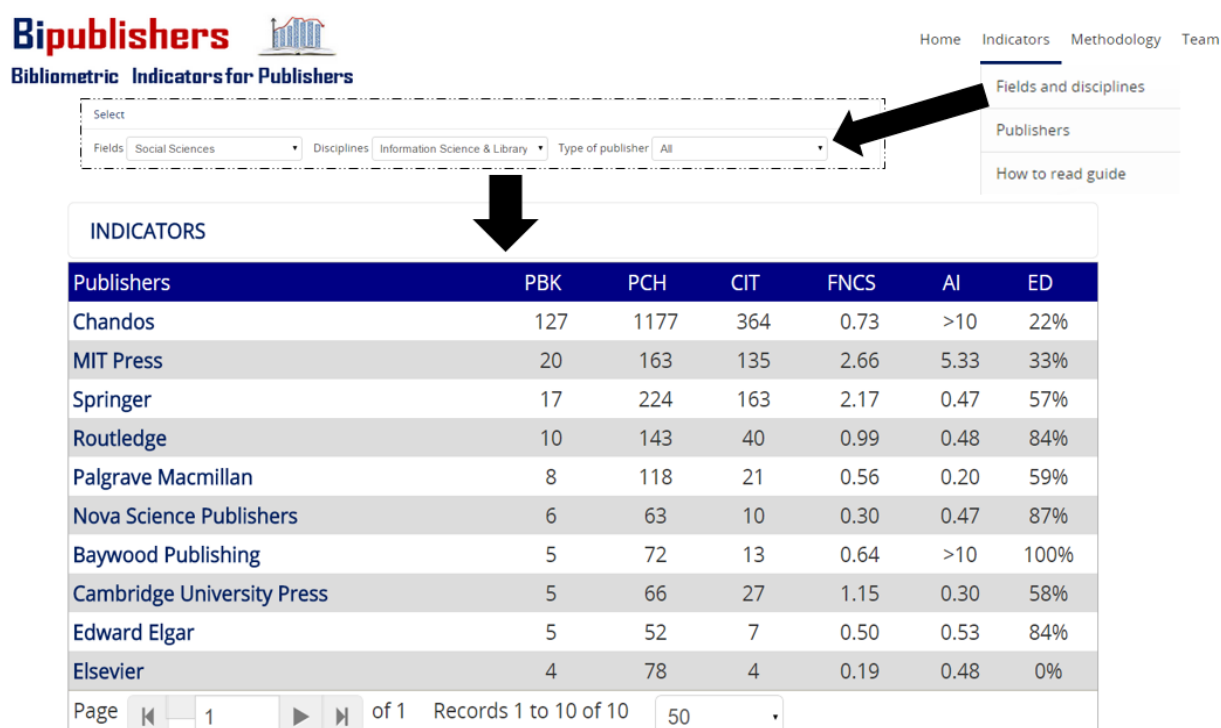

These six indicators are displayed for each publisher and by fields and disciplines. Figures 1shows how these indicators are visualised in the website. Two main entrances are available for consulting the rankings. The first one is to browse by fields and disciplines. Here the use can first select the field and then the discipline they wish to consult and can filter the results depending on the publisher type. In Figure 1 we show the ranking for the discipline of Information Science & Library Science which is included under the field of Social Sciences. As observed, rankings are sorted by default by the total number of books (PBK), however they can be resorted by the user by clicking on each header.

The other visualisation option is to look up directly for a specific publisher. Here the user can search directly for any of the publishers included in the rankings. The publisher profile page shows two tabs at the top of the page. The first tab (Data) shows basic information of the publisher (name and website). The tab normalization shows the name variants processed and included under that particular publisher, along with the city and address assigned to that given variant. Under these two tabs all fields and disciplines in which the publisher is included are displayed along with the values of the six indicators for each field or discipline. Again, results are sorted by default by the number of books.

## Main characteristics of the Bipublishers rankings

A total of 482,470 items were processed for the 2009-2013 time period. We identified 342 publishers. From this, 254 publishers are showcased. We created 42 rankings: four rankings by broad fields and 38 by disciplines. Table II offers an overview on how publishers, disciplines, citations and items are distributed among these four broad fields. As observed, Engineering & Technology is the field where less disciplines are displayed (juxst 4). However, this field is the one with the highest number of citations, showing the highest average of citation by book (5.93).

The publishers with the highest number of books indexed in the Book Citation Index are Springer (3,799 books), followed by Palgrave MacMillan (4,213) and Routledge (2,176). From the top 20 most productive publishers in the Book Citation Index (Robinson et al., 2014), only 7 are university presses, while the rest are commercial publishers. The three most productive university presses are Cambridge University Press (1,755 books), Princeton University Press (599) and University of California Press (552).

Publishers are distributed evenly in all fields except Science. Here there are fewer publishers (37) and all of them except two are commercial. Also in the field of Social Sciences there are significantly more commercial publishers (61) than university presses (23). Regarding the distribution of document types, books in Arts & Humanities have the lowest average of book chapters by book with a value of 9.8, it is followed by Social Sciences (10.7). On the other end, Science shows an average of 14.1 chapters by book while Engineering & Technology have an average of 12.0.

**Table II.** General overview of the number of publishers analysed by broad areas

| Field | No. disciplines | No Citations | No publishers | No Items | Citation average by document type |
|---|---|---|---|---|---|
| **Humanities & Arts** | 13 | 35918 | ●Commercial 38<br>●University press 41<br>●**Total: 79** | ● Books: 8864<br>● Book Chapters: 87028<br>● **Total: 95892** | ● Books 3.23<br>●Book Chapters 0.08 |
| **Social Sciences** | 14 | 59609 | ●Commercial 61<br>●University press 23<br>●**Total: 84** | ● Books: 10782<br>●Book Chapters: 114957<br>● **Total: 125739** | ● Books 4.10<br>●Book Chapters 0.13 |
| **Engineering & Technology** | 4 | 86324 | ●Commercial 37<br>●University press 38<br>●**Total: 75** | ● Books: 2820<br>● Book Chapters: 33888<br>● **Total: 36708** | ● Books 5.93<br>●Book Chapters 0.35 |
| **Science** | 12 | 28591 | ●Commercial 35<br>●University press 2<br>●**Total: 37** | ● Books: 7757<br>●Book Chapters: 109559<br>● **Total:117316** | ● Books 5.44<br>●Book Chapters 0.40 |

**Relevant publishers in Bipublisher ranking**

In Table III we include the top publishers with the largest number of books (PBK) by area with their performance indicators. As observed, there are differences on the most present publishers between Science and Engineering & Technology and Social Sciences and Humanities & Arts. While Palgrave Macmillan and Cambridge University Press are only present in the two latter fields, Elsevier and Nova Science Publishers are only present in the former. On the other hand we observe that Springer is present in all fields, however their activity index (AI) shows low values for Humanities & Arts and Social Sciences (0.29 and 0.48 respectively), while it is much higher in Engineering & Technology and Sciences (2.48 and 2.09 respectively).

**Table III.** Relevant publishers and their indicators based on four broad fields in Bipublisher

| **HUMANITIES & ARTS** | | | | | | |
|---|---|---|---|---|---|---|
| **Publisher** | **PBK** | **PCH** | **CIT** | **FNCS** | **AI** | **ED** |
| Palgrave Macmillan | 2108 | 19554 | 5772 | 0.68 | 1.42 | 38% |
| Cambridge University Press | 1004 | 8167 | 4624 | 1.48 | 1.63 | 45% |
| Routledge | 748 | 8303 | 3128 | 0.82 | 0.98 | 40% |
| Springer | 383 | 4725 | 2418 | 1.12 | 0.29 | 59% |
| Princeton University Press | 339 | 3022 | 3534 | 2.57 | 1.61 | 24% |
| **SOCIAL SCIENCES** | | | | | | |
| **Publisher** | **PBK** | **PCH** | **CIT** | **FNCS** | **AI** | **ED** |
| Palgrave MacMillan | 2680 | 26823 | 9249 | 0.68 | 1.49 | 42% |
| Routledge | 1540 | 17427 | 9077 | 0.93 | 1.65 | 45% |
| Edward Elgar | 814 | 10208 | 3434 | 0.84 | 1.91 | 62% |
| Springer | 787 | 9779 | 6734 | 1.17 | 0.48 | 67% |
| Cambridge University Press | 513 | 418 | 4373 | 2.05 | 0.68 | 49% |
| **ENGINEERING & TECHNOLOGY** | | | | | | |
| **Publisher** | **PBK** | **PCH** | **CIT** | **FNCS** | **AI** | **ED** |
| Springer | 1054 | 12139 | 14831 | 1.29 | 2.48 | 53% |
| Elsevier | 387 | 5238 | 3943 | 1.16 | 3.92 | 27% |
| Nova Science Publishers | 267 | 2665 | 954 | 0.28 | 1.79 | 77% |
| Woodhead Publishing | 192 | 2482 | 878 | 0.43 | 8.02 | 73% |
| Artech House | 142 | 1759 | 676 | 0.55 | 8.08 | 30% |
| **SCIENCE** | | | | | | |
| **Publisher** | **PBK** | **PCH** | **CIT** | **FNCS** | **AI** | **ED** |
| Springer | 2446 | 40396 | 37013 | 1.17 | 2.09 | 73% |
| Nova Science Publishers | 961 | 10711 | 3079 | 0.24 | 2.34 | 77% |
| Elsevier | 538 | 10711 | 7787 | 1.28 | 1.98 | 56% |
| Cambridge University Press | 417 | 5036 | 5728 | 1.61 | 0.77 | 44% |
| Routledge | 361 | 4964 | 1948 | 0.47 | 0.54 | 43% |

Regarding their impact, the only two university presses included in the top 5 (Cambridge University Press in Humanities & Arts, Social Sciences and Science; and Princeton University Press present in Humanities & Arts) present always values above 1 according to their normalized citation impact (FNCS), highlighting the impact of their publications. Regarding the commercial publishers, Springer and Elsevier are the only ones that show values above 1, while the rest underperform according to their FNCS.

## Methodological problems

In this paper we describe an initiative to create rankings for university presses and commercial publishers based on citation data. The data source selected was the Book Citation Index. Books and book chapters are document types of a very different nature to that to which bibliometricians are accustomed to deal with (Zuccala et al., 2014). This raises new challenges different to those raised when dealing with journal publications. In this section we will describe the main challenges observed on the development of publisher rankings.

*1. Names variants*

Thomson Reuters provides a masterlist of 499 publishers[1], however, after analyzing it we detected many errors, leading us to elaborate our own normalization process. For example, 15 name variants were detected in the case of Elsevier. Also, decisions had to be made on how such normalization process was undertaken. Unlike with journals, publishers may belong to bigger publisher corporations or may have different divisions. One should consider if a publisher ranking should include all divisions of a single publisher, maintain as separate publishers those belonging to the same corporation, or normalise to the highest level found. Here we opted for this last option; however the rationale followed for opting for one option or the other is questionable no matter which option is taken.

---

[1] http://wokinfo.com/mbl/publishers/

*2. Publisher clusters and corporations*

Following the case of Elsevier and following the criteria described above, we have included within this corporation, publishers such as Pergamon, Academic Press or North Holland, all of them belonging to Elsevier. Because the publisher market is highly unstable and subjected to continuous changes, such changes threaten the stability of the rankings and comparisons between updates. The latest change in this sense affects directly to the largest publishers included in the Book Citation Index: MacMillan and Springer, merged recently (Schweizer, 2015).

Another example we found was the case of Willan Publ, which was bought by Taylor & Francis. More difficult is taking this type of decisions when the sale is made within the study time period. This is the case of AK Peters, which was acquired by CRC Press in 2010. Finally, we must note that this issue presents serious challenges as not always the dependence relation is clear.

*3. Construction of fields and disciplines*

As mentioned before, the construction of fields and disciplines has been done by aggregating subject categories from the Book Citation Index. This is a relatively common practice in bibliometric studies when working with journal publications. In that case, journals are assigned to one or more categories. Following this line of thought, one could suggest that publishers should be assigned to categories. However, and following a more reasonable (but also less transparent) approach, every book is assigned to one or more categories. It would be of interest to better learn according to which criteria does the Book Citation Index classifies books. Also, the proposed aggregation in this paper could be questioned, hence we highlight the need to explore further alternatives.

*4. Publication types: Serials vs. books.*

A serious limitation of the Book Citation Index, is the inclusion of serials such as proceedings in the database (Torres-Salinas et al., 2013). In order to use this database for bibliometric purposes, this type of output must be removed before the analysis. In this sense, all records labelled as serials were removed from our data set; that is, records belonging to the publisher Annual Reviews (as suggested by Torres-Salinas et al., 2013).

*5. Publisher coverage*

An important limitation when analysing the output of publishers in the Book Citation Index is that we do not know what the extent of its coverage by publisher is. Do they include all books published by a publisher? Do they index only some of them? After a quick look, it seems that this latter option is the most plausible. However, further research is needed to confirm this point.

## Concluding remarks and further developments

In this paper we present the first results of the Bibliometric Indicators for Publishers project (also known as BiPublishers). This project intends to analyse the possibility of developing bibliometric indicators for scientific and academic publishers, and is the first bibliometric ranking of such characteristics. It is an on-going project currently based on data from the Book Citation Index. This means that the results displayed inherit all the shortcomings of the database. Among other limitations we highlight the bias towards English language and concentration of publishers. We discuss the main challenges that developing a bibliometric ranking for publishers entail, such as normalising publisher names, dealing with publisher merging, the construction of fields and rankings, the exclusion of certain publication types included in the Book Citation Index, as well as uncertainties as to the coverage by publisher of this database.

In order to analyse the validity of our results as well as to explore other data sets, we expect to include in the future other data sources (i.e., Scopus) as well as develop and include new bibliometric indicators that can better capture other characteristics of publishers. For instance, we suggest analysing the role of book series within publishers. In conclusion, we believe that the emergence over the last years of new citation databases including books and book chapters should encourage the bibliometric community to deepen on new venues to analyse the research impact of these long neglected document types.